\documentstyle[12pt]{article}

\begin{document}
\begin{titlepage}
\begin{center}
\vspace{1cm}
\hfill
\vbox{
    \halign{#\hfil         \cr
           yy.mm.nnn \cr
           IPM/P-99/8\cr
           Mar. 1999\cr} }
      
\vskip 1cm
{\Large \bf
Brane Mechanics}
\vskip 0.5cm
{\bf R. Abbaspur\footnote{e-mail:abbaspur@netware2.ipm.ac.ir} 
}\\ 
\vskip .25in
{\em
Institute for Studies in Theoretical Physics and Mathematics,  \\
P.O. Box 19395-5531,  Tehran,  Iran.\\
Department of Physics,  Sharif University of Technology, \\
P. O. Box 19365-9161,  Tehran,  Iran.}
\end{center}
\vskip 0.5cm

\begin{abstract}
In this paper, we investigate about two physically distinct
classes of the `one-dimensional'
worldvolume solutions describing the status of
an arbitrary brane in the presence of another arbitrary  
(flat) brane which supplies the required supergravity background.
 One of these classes concerns with a relative (transverse) motion of 
two parallel flat branes, while the other class
is related to a static configuration in which one of the branes is
flat and the other is curved as a cylindrical hypersurface. 
Global symmetries of the worldvolume theory are used to show that both  
types of these solutions are described by some sort of
`planar orbits' which are specified by their `energy' and `angular
momentum' $(E,l)$ parameters. We find that
various phases of the `motion' along these orbits, for different
values of $(E,l)$, are easily  deduced from the curve of an  
$E$-dependent function of the relative `distance' between the two branes,
 which is somehow related to their mutual
`effective potential'.

\end{abstract}

\end{titlepage}\newpage

\def\ra{\rightarrow}
\def\={:=}
\def\p{\partial}

\def\a{\alpha}
\def\b{\beta}
\def\c{\gamma}
\def\d{\delta}
\def\e{\epsilon}
\def\ve{\varepsilon}
\def\g{\gamma}
\def\v{\upsilon}
\def\th{\theta}
\def\l{\lambda}
\def\m{\mu}
\def\n{\nu}
\def\o{\omega}
\def\x{\xi}
\def\r{\rho}
\def\s{\sigma}
\def\t{\tau}
\def\f{\phi}

\def\G{\Gamma}
\def\D{\Delta}
\def\L{\Lambda}
\def\O{\Omega}

\def\cA{{\cal  A}}
\def\cB{{\cal  B}}
\def\cF{{\cal  F}}
\def\cJ{{\cal  J}}
\def\cL{{\cal  L}}

\def\Db{{\bar D}}
\def\dt{{\tilde d}}

\def\st{sin\theta}
\def\sbt{sin^2\theta}
\def\ct{cos\theta}
\def\cbt{cos^2\theta}
\def\ea{\eta^\alpha}
\def\bX{{\bf X}}

\def\ft{\footnote}
\def\nn{\nonumber}
\def\be{\begin{equation}}
\def\ee{\end{equation}}
\def\bea{\begin{eqnarray}}
\def\eea{\end{eqnarray}}
\def\np{\newpage}

\def\leg{<=>}
\def\ts{\ \ \ \ \ \  ''}

\section{Introduction}
The description of the composite systems
of branes \cite{-2,-1,0} in terms of
the source and probe (S,P) configurations has proved itself as a useful
technique for investigating about some of the features of
 the intersecting brane configurations \cite{1,2,3}.
 As has been analyzed in \cite{1}, describing
a system of branes
in this manner, one can  
find a subset of the `field constraints' (the no-force conditions)
that fix the supergravity solution for
a `marginal' configuration of
`flat' intersecting or overlapping branes. Moreover,
by this analysis several
`algebraic constraints',
 arising naturally as a the `consistency conditions' 
 in the study of the supergravity field equations,
 acquire interpretations as the
 conditions for balancing several internal long-range forces
 against each other.
Equivalently, one can interpret them
as the balancing conditions for the forces acting 
between the whole of the brane system and external brane probes of 
different types  and orientations similar to those of the
brane system \cite{1}.
A similar approach in systems of branes at angles reveals that the
only {\it marginal} configurations with flat branes at (non-trivial) angles
are those constructed from  
similar branes with two angles in a subgroup of $SU(2)$ \cite{4}. 
From the viewpoint of the supergravity theory, these configurations 
preserve 1/4 of the spacetime SUSY \cite{5,6}. There exist, however, 
other configurations of intersecting branes making more than two angles,
but still preserve some fraction of SUSY \cite{5,7}. 
These configurations are stable 
only when certain relations (which embody the above $SU(2)$ relation) 
hold between their angles \cite{7,7*}. 
Such configurations, are found in \cite{4} that,  
can not be made up of a set of flat p-branes. 
Indeed, they consist of a set of 
curved p-branes which only asymptotically  
look like the flat intersecting (overlapping) branes with definite
angles. This agrees with a result of \cite{8} that the supergravity 
solutions for the {\it non-orthogonal} intersecting branes of the 
distributed type at {\it non} $SU(2)$ angles are not realizable in terms
of a set of harmonic functions. However, the existence of such `curved 
intersections' raises the question of finding the 
corresponding worldvolume  solutions. This generally defines a difficult 
problem which involves solving simultaneously the coupled set of the
supergravity and the worldvolumes field equations. As everybody knows,
even the much simpler but analogous problem for two charged particles in 
a flat spacetime has not an exact solution. In a limiting case,
however, the problem is simplified
and can be handled using the S-P description. This
is the viewpoint that we shall follow in this paper to study a special class
of the curved configurations
 which we name as the `mono-dimensional' solutions.
 The organization of the paper
 is as follows: The general set up
of the problem together with the 
required notations are given in section 2. 
Then in section 3, the symmetries of the worldvolume theory and
its Noether currents, in connection to a class of its solutions named 
as the `orbits',  are studied. Sections 4,5,6 are devoted to a detailed 
analysis of the orbits in three different cases of the source and probe 
dimensions on the basis of the conservation equations given in section 3.  
The paper ends with some concluding remarks in section 7. 

\section{Some general remarks} 
The basic principle in the analysis of any system of (S,P)-branes 
(as the source and probe respectively) is that, while the worldvolume
geometry of the P-brane is essentially
determined by the fields that originate from
the S-brane, the latter is not influenced in any way by the
presence of the former. In analogy to classical mechanics, such a condition
corresponds to a limiting case in which one of the branes, labeled by S, has a
considerable mass relative to the other brane which is labeled by P.
In the case of BPS (1/2 SUSY) branes, one knows that each of the branes has
a mass and charge in proportion to its tension $T_s$ or $T_p$ \cite{2}; so
 in suitable units one must require that $T_s\gg T_p$.
There are other occasions, however, that while this condition
on tensions is not necessarily
 satisfied, the description as a (S,P) system still is reliable.
An interesting example of such occasions 
is  a marginally bound BPS configuration of two flat p-branes
which are characterized by certain no-force conditions
among themselves \cite{1,2,9}.
In general, a BPS-brane with a small (extrinsic) curvature
in the background of another such flat brane,
both together,  can be treated (in this same order)
as a pair of (P,S)-branes. Another application concerns the near to the
 intersection points of two p-branes
where one can neglect the self-interactions of each brane
in comparison to its interactions with the other brane.
In usual, a flat p-brane of the BPS
type is a classically stable object, which obeys certain no-force conditions
preventing its collapse due to its self-gravitational forces. When
the p-brane is curved, the
self-interaction forces counterbalance
each other and the oscillations initiate.
 In a S-P description, however, we deal with a limiting case in
which all of the `self-oscillations' of the P-brane, as well as its curving 
effects on the  geometry of the S-brane,  are totally negligible. 
As a result, the analysis for a S-P
configuration with a pair of BPS-branes proceeds via studying
the worldvolume (DBI+WZ) action of the P-brane, in which the background fields
are replaced by the fields associated to a flat and fixed S-brane. 
In general, the analysis depends on the `types' of  the two branes
which-as far as concerns to us-
 are specified by their dimensions, `electric'  or `magnetic' nature
and couplings to the dilaton field.
For the purpose of this paper we restrict our study
to the case  with both electric-type (S,P)-branes and
distinguish between several
cases by the relative status of the
worldvolume dimensions $(d_s,d_p)$ of these two branes.\\ 
To begin with, we recall that the general supergravity
background for a (BPS) $(d-1)$-brane source consists of \cite{2,3}
\bea
&&ds^2=H^{d/\Db}(H^{-1}{\eta}_{\m\n}dx^{\m}dx^{\n} +{\d}_{mn}dy^mdy^n)\nn\\
&&e^\f =H^{\a}\nn\\
&&\cA =H^{-1}{\L}_{\m =0}^{d-1}dx^{\m}
\label{1}
\eea

Here $x^\m$ and $y^m$ are parallel and transverse coordinates of the S-brane,
and $H$ is a harmonic function:
\bea
          H(y)=\left\{ \begin{array}{lll}
          1+Q/ |y|^{\dt}\;\;\;\; &,&\;\;\;\; \dt\neq 0\\
          -Q\ln |y|\;\;\;\; &,&\;\;\;\; \dt =0 
                          \end{array}
                          \right.
\label{2} \eea
where $|y|^2=y^my^m$ and $Q$ is 
(proportional to) the form-field charge of the brane.
All other notations are those of \cite{1}, in particular: $\Db\= D-2 , \ \
\dt\= \Db -d$, D is the spacetime dimension and $\a$ is a measure of coupling
between $\f$ and $\cA$ in supergravity, which is related to $d$ as: \cite{2,3}
\be
{\a}^2(d)=1-d\dt /2\Db
\label{a}\ee
On the other hand the worldvolume action for a
$(d-1)$-brane probe (having equal mass and charge in units with 
$16\pi G_D=1$) has 
 the general form \cite{2,9*}:
\be
S_d=-T_d\int d^d\xi e^{-\a\f} {det}^{1/2}(g^*_{ab})+T_d\int{\cA}^*_{(d)} 
\label{3}
\ee
where $g^*_{ab}$ and ${\cA}^*$ are pull-backs of the spacetime metric 
and the $d$-form potential, on the $d$-volume of the probe, respectively.
Depending on that $d_s=d_p$ or $d_s\neq d_p$,
the last (WZ) term has a non-vanishing or vanishing contribution to $S_d$. 
 This action describes the dynamics of the embedding 
coordinates $X^M=X^M(\xi)$ through the dependences of $\f , g^*_{ab} ,  
{\cA}^*$ on these coordinates.
In the following sections,however, we use a subset
$(x^a)$ of the spacetime coordinates $(x^M)$ (parallel or 
perpendicular to the source) to parameterize the worldvolume of the P-brane
and use the remaining subset $(y^A)$ as the embedding coordinates
[so, $(x^a, y^A)\equiv (x^{\m},y^m)$]. The above action then reads 
\be
S_d=T_d\int d^dx\cL (Y^A,{\p}_aY^A)
\label{4} 
\ee
where the form of the embedding Lagrangian ($\cL$) changes depending on the
relative status of $d_s$ and $d_p$ (see below).

\section{Symmetries of $\cL$ and the orbits}
Without going to the detailed expression
of $\cL$ in the specific cases, the
flat geometry of the S-brane a priori implies that this Lagrangian must possess a
number of obvious global symmetries.
These symmetries include the translational and
rotational invariances along the worldvolume of the S-brane as well as the
rotational invariance along its
transverse directions which are respectively defined as:
\bea
\d x^{\m}={\ve}^{\m}\;\;\; &,&\;\;\;\d y^m =0\;\;\;\;\;\; 
\label{5}\\
\d x^{\m}={\l}^{\m}_{\n}x^{\n}\; &,&\;\;\; \d y^m =0\;\;\;\;\;\;\;\;\;
({\l}_{\m\n}+{\l}_{\n\m}=0)
\label{6}\\
\d x^{\m}=0\;\;\; &,&\;\;\;\d y^m={\o}_{mn}y^n \;\;\;\;\;\; 
({\o}_{mn}+{\o}_{nm}=0)
\label{7}\eea

Though all of these symmetries have natural interpretations as the spacetime
(bulk) symmetries, in the worldvolume theory some of them are realized as
pure internal symmetries changing $Y^A(x^a)$'s but not affecting on $x^a$.
The three cases with $d_s\leg d_p$ are common in the above symmetries,
though they are different in the way that one defines $(x^a,y^A)$. 
In each of the above cases they are defined as follows:
\bea
&&d_s=d_p\;\;\;\; :\;\;\;\;x^{\m}=x^a\;\;\;\; ,\;\;\;\;\;\ \ \ y^m=y^A\nn\\
&&d_s>d_p\;\;\;\; :\;\;\;\;x^{\m}=(x^a,y^i)\;\;\;\; ,\;\;\;\; y^m=y^A\nn\\
&&d_s<d_p\;\;\;\; :\;\;\;\;x^a=(x^{\m},y^i)\;\;\;\; ,\;\;\;\; y^m=(y^A,y^i)
\label{9}\eea
Here $(y^i)$ represents a subset of the coordinates parallel (perpendicular)
to the S-brane which we shall take them  as a part of the parameterizing
(embedding) coordinates, in the case with $d_s>d_p \ (d_s<d_p)$. 
The corresponding Noether currents for each of these cases
are then constructed using the formula \cite{10}:
\be
{\cJ}^a={{\p\cL}\over{\p Y^A_{,a}}}(\d Y^A-\d x^b{\p}_bY^A)+\cL \d x^a
\label{8}
\ee
The conservation equations of the Noether currents are useful tools
in the analysis of the classical solutions of 
the worldvolume theory (\ref{4}) for a particular class of them
depending only on a single space or time coordinate. For the reason that
becomes clear, we refer to such solutions as the `orbits' of the P-brane and 
denote by $x$ the space or time coordinate that parameterizes an orbit.
For such a configuration, the action (\ref{4})
for the worldvolume field theory reduces
to a mechanical type action of the form
\be
S=\int dxL(Y^A,{\dot Y}^A)
\label{10}\ee
where $Y^A=Y^A(x) ,\ {\dot Y}^A\=dY^A/dx$, and $L(Y^A,{\dot Y}^A)$ (up to a
constant factor) is the same as $\cL (Y^A,{\p}_aY^A)$ 
in which all derivatives of $Y^A$ other 
than ${\dot Y}^A$ are set equal to zero. Depending on that $x$ is a space-
or time-like coordinate the resulting solution has a different interpretation.
When $x$ is space-like, $Y^A(x)$ describes a cylindrical static configuration
of the P-brane, whose all directions except one are parallel to the S-brane.
When $x$ is time-like, $Y^A(x)$ describes the path of the motion for a P-brane
remaining parallel to a S-brane and moving transverse to it. In the subspace
transverse to all of the $x^a$ directions, the configuration (of the P-brane)
defined by $Y^A=Y^A(x)$ is projected into a curve which looks like the
orbit of a point particle moving within a spherically symmetric potential. 
The adjective `space- (time-)like' for the orbits
then signifies that the corresponding curve in the subspace of $(x,y^A)$ is 
space- (time-)like, meaning that $x$ itself is a space- (time-)like coordinate.
For both classes of such {\it mono-dimensional} solutions, the equations of
motion of (\ref{10}) can be more easily integrated using a set of conservation
laws corresponding to the symmetries given by
(\ref{5})-(\ref{7}). Generically, for an orbit depending on $x$, a
conservation low as ${\p}_a{\cJ}^a=0$ yields
\be
{{d{\cJ}^x}\over{dx}}=0
\label{11}\ee
which identifies the component ${\cJ}^x$ of the current as a conserved
charge along the orbit. We will see in the following sections that these
conserved charges provide the sufficient information required for integrating
the equations of motion for $Y^A(x)$. Then we observe that the orbits are
effectively characterized by the 2-dimensional motion of a point particle
in the Newtonian mechanics whose different phases of motion are determined
by an effective potential depending on two (energy and angular momentum)
parameters. The analysis then proceeds via a set of (phase) diagrams depending
only on the energy parameter, and reveals a variety of regimes (phases of the
effective motion) for orbits with various parameters.
We study the three cases with $d_s<=>d_p$ through separate analyses in the
next three sections.

\section{The S-P system with $d_s=d_p$}
In this case the $1^{st}$ row of (\ref{9}) defines $(x^a,y^A)$ and the
Lagrangian $\cL$, using eqs.(\ref{1})-(\ref{3}), for a general
configuration is written as
\be
\cL=-H^{-1}(Y)[det^{1/2}({\d}^{\m}_{\n}+H(Y){\p}^{\m}Y^m{\p}_{\n}Y^m)-\eta]
\label{12}\ee
where $\eta =+1(-1)$ for a brane-brane (brane-anti-brane) S-P
configuration (note that in this case the two branes `see' the form-fields
of each other via a force depending on their relative orientation; the
sign of $\eta$ reflects this orientation). The conserved currents in this
case consist of:
\bea
T^{\m}_{\ \n}={{\p\cL}\over 
{\p Y^m_{\ ,\m}}}Y^m_{\ ,\n}-{\d}^{\m}_{\n}\cL \;\;\;\; 
&,&\;\;\;\; {\p}_{\m}T^{\m\n}=0
\label{13}\\
M^{\m\n\r}=x^{\r}T^{\m\n}-x^{\n}T^{\m\r}\;\;\;\;
&,&\;\;\;\;{\p}_{\r}M^{\m\n\r}=0
\label{14}\\
J^{\m}_{mn}=Y^m{{\p\cL}\over 
{\p Y^n_{\ ,\m}}}-Y^n{{\p\cL}\over {\p Y^m_{\ ,\m}}}
\;\;\;\; &,&\;\;\;\;{\p}_{\m}J^{\m}_{mn}=0
\label{15}\eea
which in correspondence to the symmetries (\ref{5})-(\ref{7}) have to be
identified as the energy-momentum, `longitudinal' and `transverse' angular
momentum tensors respectively. Since the conservation of $M^{\m\n\r}$
naturally follows from that of $T^{\m\n}$, 
only the eqs.(\ref{13}) and (\ref{15})
are taken as the informative parts of the conservation laws for this analysis.
Going to the mono-dimensional solutions, the Lagrangian (\ref{12}) becomes
\be
L=-H^{-1}[(1+\ve H{\dot Y}^2)^{1/2}-\eta]
\label{16}\ee
where $\ve =+1(-1)$, if $x$ is a space- (time-)like coordinate and
${\dot Y}^2\={\dot Y}^m{\dot Y}^m$. The corresponding conserved charges,
according to eqs.(\ref{11}),(\ref{13}),(\ref{15}), are
\bea
E&=&{\dot Y}^m{{\p L}\over{\p {\dot Y}^m}}-L=-H^{-1}\left (\eta -
{1\over{\sqrt{1+\ve H{\dot Y}^2}}} \right )
\label{17}\\
J^{mn}&=&{\dot Y}^m{{\p L}\over{\p {\dot Y}^n}}-
{\dot Y}^n{{\p L}\over{\p {\dot Y}^m}}=-{{\ve (Y^m{\dot Y}^n-Y^n{\dot Y}^m)}
\over {\sqrt{1+\ve H{\dot Y}^2}}}
\label{18}\eea
For a time-like orbit ($\ve =-1$), $E$ and $J^{mn}$ have obvious
interpretations as the energy
\footnote{This is indeed the {\it binding} energy of the configuration
in units of the source tension $T_s$, which excludes the rest
masses of (S,P)-branes.}
 and the (transverse) angular momentum tensor
(of the P-brane) respectively. For a space-like orbit ($\ve =+1$), however, 
they have to be interpreted as the flows of the respective quantities along
the orbit ($x$) direction \footnote{It is easy to check that in both cases
$T^{\m}_{\ \n}$ is a diagonal matrix, with $T^x_{\ x}=E$ and its all other diagonal
components are equal to $(-L)$. So, for $\ve =-1$, $E$ is the energy while
for $\ve =+1$ it is a component of the (anisotropic) pressure along the $x$
direction.}.\\
As in the ordinary 3D particle mechanics, one expects that the angular
momentum conservation restricts the dimension of the orbits (hyper)planes.
Such a property is in fact a result of the equation(s) obtained from
eq.(\ref{18}) by eliminating the velocities; that is
\be
J^{mn}Y^p+J^{np}Y^m+J^{pm}Y^n=0
\label{19}\ee
where distinguished equations are 
obtained for distinct values of $m,n,p=1,...,M$
with $M\={\dt}_s+2$. So, the orbit $Y^m=Y^m(x)$ lies at the common
intersection of a set of 
$(M-1)$-dimensional hyperplanes defining the
`orbit plane'. The (maximum) number of independent equations of the form
eq.(\ref{19}) thus fixes the (maximum) dimension of the orbit plane.
For $M>3$ one can show that this plane, just as in the Newtonian particle
mechanics, is a 2-dimensional plane, provided
\footnote{For $M=1,2,3$, one have
respectively 0,1,3 non-zero components for $J^{mn}$ and a relation of the
type of eq.(\ref{20}) is not required.}
\be
J^{12}J^{34}+J^{13}J^{42}+J^{14}J^{23}=0, \;\;\;etc.
\label{20}\ee
Further, when (all) $J^{mn}=0$, the orbit plane degenerates to a straight line
or a single point. By rotating the $y^m$ coordinates, so that the orbit plane
coincides with the $y^1y^2$-plane, all $J^{mn}$'s except $J^{12}$ are 
transformed to zero (obviously, such $J^{mn}$'s obey eq.(\ref{20})). Using 
the polar coordinates $(r,\th )$ in the $y^1y^2$-plane, the conservation  
laws (\ref{17}),(\ref{18}) will read
\bea
{1\over{H(r)}}\left (-\eta+
{1\over{\sqrt{1+\ve H(r)({\dot r}^2+r^2{\dot\th}^2)}}} \right )&=&E
\label{21}\\
{{r^2{\dot\th}}\over{\sqrt{1+\ve H(r)({\dot r}^2+r^2{\dot\th}^2)}}}&=&l
\label{22}\eea
where $H(r)$ is the same as $H(y)$ (eq.(\ref{2})) with $r\equiv |y|$, and 
$l\=-\ve J^{12}$ is the angular momentum parameter. 
Solving these two eqs.for $(\dot r, \dot{\th})$, 
one obtains
\bea
&&{\dot r}^2=U(r)\=
{{-2\ve\eta E-\ve E^2H-l^2/r^2}\over{(\eta +EH)^2}}
\label{23}\\
&&\dot{\th}=\o (r)\={{l/r^2}\over {\eta +EH(r)}}
\label{24}\eea
These two equations, in principle, determine the orbit up to the integrations 
\bea
&&x=\pm {\int}^r{{ds}\over\sqrt{U(s)}}
\label{25}\\
&&\th =\pm {\int}^r ds{{{\o}(s)}\over{\sqrt{U(s)}}}
\label{26}\eea
where the choice of the $+\ (-)$ sign corresponds to 
an incoming (outgoing) effective radial motion.   
Solving  eqs. (\ref{25}),(\ref{26}) for $(r(x),\th (x))$ then specifies
the form of the projection of the P-brane as a 3D curve in the $(x,y^1,y^2)$
subspace.
  However, the shape of the orbit  in
the $y^1y^2$-plane can be directly  determined by solving a
 differential equation, obtained from
eqs.(\ref{23}),(\ref{24}), as
\be
\left ({{dr}\over{d\th}}\right )^2={{U(r)}\over{{\o}^2(r)}}=
{{r^2}\over{l^2}}[f(r)-l^2]
\label{27}\ee
where $f(r)$ is defined in terms of $H(r)$ as
\be
f(r)\=-\ve r^2[2\eta E+E^2H(r)]
\label{28}\ee
In terms of this function, $U(r)$ and $\o (r)$ (eqs.(\ref{23}),(\ref{24}))
are expressed as
\bea
&&U(r)={{E^2r^2(f(r)-l^2)}\over{(f(r)+\ve\eta Er^2)^2}}
\label{29}\\
&&\o (r)={{\ve lE}\over{f(r)+\ve\eta Er^2}}
\label{30}\eea

\subsection{The effective one-dimensional motion}
eq.(\ref{23}) is similar to an energy conservation equation for the 
1-dimensional (radial) motion of
a point particle within an external force field
derivable from the potential 
$-{1\over 2}U(r)$ and with $x$ identified as the time
variable. Except for a few examples,  
explicit integration of this equation for $r(x)$ is generally impossible. 
However, as in many other dynamical systems, the qualitative behavior of the
solution and its different phases are distinguishable from a set of 
`phase diagrams'; i.e. the curves describing the motion in the $(r,\dot r)$ 
plane. These curves are, in effect, the symmetrized version of the $U(r)$   
diagrams for different values of the $(E,l)$ parameters. However, a more 
simplified discussion of the effective
motion proceeds via studying the diagrams for
$f(r)$ which are dependent only on the $E$ parameter. The expression
of eq.(\ref{29}) for $U(r)$ 
shows that the allowed region of the `effective radial motion' are specified 
by the values of $r$ satisfying the inequality 
\be
U(r)\geq 0\ \ \ \ \ \ \equiv \ \ \ \ \ \ f(r)\geq l^2
\label{31}
\ee
Within every connected region of $r$ with this property,
the `effective particle' moves in a definite direction (depending on the sign
in eq.(\ref{25})) under the radial force 
${1\over 2}U'(r)$, until it reaches to an
end-point of the interval, where its velocity vanishes and (provided the
force is still non-vanishing) reverses its direction. Obviously, such
`turning points' are defined by
\be
f(r_0)=l^2\ \ \ ,\ \ \ f'(r_0)\neq 0
\label{turn}\ee

In addition, the positivity of the square root in in eq.(\ref{21}) requires
that
\be
\eta +EH(r)>0
\label{eta}\ee
By eq.(\ref{24}), this result implies that the angular `velocity'
$\o (r)$ has a definite sign (depending on $l$), which is an
indication of a {\it monotonic} angular `motion' around the origin.
 By eqs.(\ref{2}) and (\ref{28}), the explicit expression for $f(r)$ is
\bea
          f(r)=\left\{ \begin{array}{lll}
          \ve (ar^2-{b/{r^{M-4}}}) \;\;\;\; &,&\;\;\;\; M\neq 2\\
          \ve\left (cr^2+br^2 \ln r\right ) \;\;\;\; &,&\;\;\;\; M =2 
                          \end{array}
                          \right.
\label{32} \eea
where the various parameters are defined as\footnote{Note to the
dimensionality of the parameters as: \\
$E\sim (length)^0\ \ \,\ \ \ l\sim (length)^1\ \ \,\ \ \ Q\sim (length)^{M-2}$
}:  
\be
a\= -2\eta E-E^2\ \ \ \ ,\ \ \ \ b\= E^2Q\ \ \ \ ,\ \ \ \ c\= -2\eta E
\label{33}
\ee
For each specific situation (i.e. a case with a specific $M, \ve ,\eta  $),
 fixing the energy $E$ of the P-brane specifies the shape of
the graph for $f(r)$. Changing $l^2$, as the height of a horizontal line on 
this graph, then according to eq.(\ref{31}), specifies various 
allowed regimes of the effective motion. 
Evidently, in all cases that $f(r)\equiv 0\ (\forall r\geq 0$), one
necessarily has $l=0$ and in such a case any {\it constant}
values of $(r,\th )$ specify a solution of the equations of motion.
Such a solution, which is the same as a multi-center solution for
two parallel branes, can be formed at any arbitrary separation of the
two branes thereby defining  a {\it marginal} configuration \cite{3,0}. By
eqs.(\ref{32}),(\ref{33}), such a configuration is only achieved for
$E=0$, i.e. for a zero binding energy. This is the only configuration of
two similar branes in the desired category
 which is marginal and preserves a 1/2 SUSY fraction.

\subsection{Qualitative description of the orbits}
The following qualitative descriptions are deduced from the inspection
of the the graphs of $f(r)$ in Fig.(1), which are sketched for the
space-like orbits, and those for the time-like orbits which are obtained
by reversing these graphs (not shown). The parameter $a$ in these diagrams
is related to the energy parameter $E$ as in the eq.(\ref{33}).\\

{\it Space-like orbits ($\ve =+1$)}\\
For all the transverse dimensions $M\geq 3$, these orbits share the common
feature that, for all values of $l$, they have extensions to the infinite
region $r\rightarrow\infty$ and pass through a minimum distance $(r_0)$
relative to the source. They are all symmetric around their
minimum points and make a finite angle $(\c)$ with respect to the $x$
direction (see below). All of these `solitons' are stable only for those
combinations of $(E,\eta )$ that leave $a>0$. For $M=1$, which normally has
$l=0$, the orbit crosses the source (at $r=0$) and passes through a maximum
distance $(r_0)$ and oscillates periodically along the $x$ direction. The
stability in this case is also achieved for $a>0$. The $M=2$ case is
distinguished by the fact that the coordinate $r$ possesses a natural bound
$r\leq 1$ so that $H(r)\geq 0$ and the transverse part of the metric remains
positive. The corresponding orbits then oscillate between $r_0\leq r\leq 1$,
and are stable provided $c>0$ (so that $r_0<1$).\\

{\it Time-like orbits ($\ve =-1$)}\\
Unlike the space-like case, time-like 
diagrams in all of the three branches with
$a\leg0$ possess allowable regimes of `motion'. Possible regimes are:\\

$M\geq5\ ,\ a\geq 0:\ $ For all $l^2\geq 0$ the orbit is bound between
$0\leq r\leq r_0$, though for $a=0\ ,\ l=0$ it extends to infinity:
$0\leq r<\infty$.\\
$M\geq 5\ , \ a<0:\ $ There is a critical value $l_0^2$, so that the orbits with
$l^2<l_0^2$ range between $0\leq r<\infty$, while for $l^2\geq l_0^2$ two
different regimes in: $0\leq r\leq r_1$ and $r\geq r_2$ with the same amounts
of $(E,l)$ are equally possible.\\
$M=4\ ,\ a>0:\ $ For each $l^2\leq l_0^2$, the orbit is bound between
 $0\leq r\leq r_0$. Larger values of $l^2$ are not allowable.\\
$M=4\ ,\ a<0:\ $ For each $l^2\leq l_0^2$ 
the orbit ranges in $0\leq r<\infty$,
while for $l^2>l_0^2$ it ranges in $r\geq r_0$.\\
$M=4\ ,\ a=0:\ $ For all $l^2<l_0^2$, 
the orbit is a spiral between $0<r<\infty$.
For the critical value $l^2=l_0^2$, it becomes a circle of {\it arbitrary}
radius which is traversed with a constant angular velocity. \\
$M=3\ ,\ a>0:\ $ For all $l^2<l_0^2$ , the orbit is bound between $r_1<r<r_2$.
Just for $l^2=l_0^2$, the two bounds coincide and the orbit becomes a circle
with an {\it energy dependent} radius and angular velocity. \\
$M=3\ ,\ a\leq 0:\ $ For $l\neq 0$, the orbit ranges in $r\geq r_0$, while for
$l=0$ it passes through $r=0$.\\
$M=2\ ,\ a\leg0:\ $ This is the analogue of the $M=3\ ,\ a> 0$ case above.\\
$M=1\ ,\ a\leg0:\ $ This is the analogue of the $M=3\ ,\ a\leq 0$ case above,
with the difference that when $a>0$ the orbit never passes the origin even
for $l=0$.
\subsection{Quantitative features}
\subsubsection{The infinite orbits}
As the above description shows, (for both $\ve =\pm 1$ cases) there are many
regimes in which the orbit extends to the infinite region $r\ra\infty$. The
asymptotic behavior of such orbits is deduced from the study of the asymptotic
behaviors of $U(r)$ and $\o (r)$. For the present purpose, we need only to
the limiting values, which according to eqs.(\ref{29}),(\ref{30}),(\ref{32}),
are 
\bea
         r\ra\infty \;\;\;\;\left\{ \begin{array}{l}
          U(r)\ra U_0\\
          \o (r)\ra 0 
                          \end{array}
                          \right.
\label{34} \eea
where $U_0=U_0(E)$ is
\bea
          U_0=\left\{ \begin{array}{ccc}
          \ve \left ({1\over{(E+\eta )^2}}-1\right ) 
          \;\;\;\; &,&\;\;\;\; M\geq 3\\
          0 \;\;\;\; &,&\;\;\;\; M =1 
                          \end{array}
                          \right.
\label{35} \eea
 These limiting values,
according to the differential eqs.(\ref{23}),(\ref{24}), correspond to
straight line asymptotic orbits, in the $\th ={\th}_0$ planes,  with a slope
$dr/dx=\pm \sqrt{U_0}$. This means that, in all the cases with $U_0\geq 0$,
the S-P system asymptotically looks like a pair of {\it flat}
$(d-1)$-branes, with a relative rotation (boost) in one direction
 with an angle $\c$ (a velocity $v$) satisfying 
\bea
          U_0=\left\{ \begin{array}{lll}
          tan^2\c \;\;\;\; &,&\;\;\;\; \ve =+1\\
           v^2 \;\;\;\; &,&\;\;\;\; \ve =-1 
                          \end{array}
                          \right.
\label{36} \eea

Consequently, the energy-dependences of $\c$ and $v$ are determined by
eq.(\ref{35}). For $M=1$, this indicates that all values of $E$ yield to a
{\it parallel static} configuration of the S$\&$P-branes at infinite
distances. For $M\geq 3$, it gives
\bea
         E+ \eta  =\left\{ \begin{array}{lll}
           |cos\c | \;\;\;\; &,&\;\;\;\; \ve =+1\\
           1/\sqrt{1-v^2} \;\;\;\; &,&\;\;\;\; \ve =-1 
                          \end{array}
                          \right. \ \    (M\geq 3)
\label{37} \eea
Accordingly, the allowed values of $E$ for these infinite orbits range as
\bea
           \left\{ \begin{array}{ccc}
          - \eta \leq E\leq 1-\eta \;\;\;\; &,&\;\;\;\; \ve =+1\\
           E\geq 1-\eta  \;\;\;\; &,&\;\;\;\; \ve =-1 
                          \end{array}
                          \right. \ \   (M\geq 3)
\label{38} \eea
For $\ve =+1$, the bound $E=1-\eta $ corresponds to an asymptotic 
parallel ($\c =0$) or anti-parallel ($\c =\pi $) S-P configuration, while 
 $E=-\eta$ corresponds
to an asymptotic orthogonal ($\c =\pi /2$) situation.
For $\ve =-1$, the bound $E=1-\eta $ corresponds to an asymptotic static
 ($v=0$) moving configuration, while $E=+\infty$ corresponds to an asymptotic
light-like ($v=1$) motion. The values of $E$ near these two extremes 
 correspond to the Newtonian and Relativistic limits respectively.

\subsubsection{The circular orbits} 
Among the graphs of Fig.(1), regimes can be distinguished, for which 
the orbits have a constant radius $r=R$. For each specific value of $E$, such
an orbit (if any) is characterized by the values of $l^2$ that touch the 
extremum points of the graph for $f(r)$, i.e.  
\be
f(R)={l_0}^2\ \ \ \ ,\ \ \ \ f'(R)=0
\label{39} \ee
Noting to the sign of the `effective force', the maximum (minimum) points of  
$f(r)$ characterize the stable (unstable) circular orbits. Among the 
space-like cases with $M\geq 2$, we see that there are no circular orbits. 
Among the time-like orbits, however, we find a special type of the circular
orbits one for each dimension $M\geq 2$. 
These consist of the following cases:\\

$M\geq 5\ ,\ a<0 \ $; {\it Unstable circular orbits}:\\   
The radius ($R$), angular momentum ($l_0$), and angular velocity (${\o}_0$) of
the orbits, by eqs.(\ref{30}),(\ref{32}),(\ref{39}), are found to be
\bea
R&=&\left ({{M-4}\over 2}{{Q}\over{1+2\eta /E}}\right )^{1\over {M-2}}\nn\\
{l_0}^2 &=& {{M-2}\over 2}  E^2QR^{4-M}\nn\\
{\o}_0 &=&{{\left [\left ({{M-2}\over{M-4}}\right )
\left (1+2{{\eta}\over E}\right )\right ]^{1/2}}\over
{R\left [\left ({{M-2}\over{M-4}}\right )
\left (1+2{{\eta}\over E}\right )-{{\eta}\over E}\right ]}}
\label{40}\eea
It is amazing that the velocity $v_0=R{\o}_0$ for these orbits turns 
to be independent of $Q$.\\

$M=4\ ,\ a=0\ $; {\it Marginally stable circular orbits}:\\
Such orbits, having {\it arbitrary radiuses}, are  special to the dimension    
$M=4$ and occur for very specific values of $(E, l)$ which yield 
$f(r)\equiv l^2$ for all values of $r$. So, for such orbits we must have 
\be 
E=-2\eta\ \ \ ,\ \ \   {l_0}^2=4Q
\label{41}\ee
However, since by the eq.(\ref{eta})
we must have $E>-\eta $, such `marginally stable' circular orbits
can only happen in the case of a brane-anti-brane system with $\eta =1$,
having the energy  $E=2$.
A circle of radius $R$ in this case is traversed with the angular velocity
\be 
{\o}_0={{2\sqrt Q}\over{2Q+R^2}}
\label{42}\ee\\

$M=3\ ,\ a>0\ $; {\it Stable circular orbits}:\\
The formulas for $R,\; {l_0}^2,\;{\o}_0$ are just given by eqs.(\ref{40}) with 
$M=3$. Again the velocity on this circle is independent of $Q$.\\

$M=2\ $; {\it Stable circular orbits}:\\
In this case , the values of $R, {l_0}^2,{\o}_0$ are 
\bea 
&&R=e^{2\eta /EQ-1/2}\nn\\
&&{l_0}^2=1/2E^2QR^2\nn\\
&&{\o}_0={{\sqrt{Q/2}}\over{R(Q/2-\eta /E)}}
\label{43}\eea
Despite the $M\geq 5$ or $M=3$ cases, the velocity on these circles does 
depend on $Q$. Because of the natural bound $r\leq 1$ in $M=2$, for these 
orbits to be realized we must have $4\eta /EQ<1$. 

\subsubsection{The orbits crossing the origin}
Amongst the various regimes, distinguishable from Fig.(1), there are
cases where the orbit can approach, or pass through, the origin ($r=0$).   
Such regimes (orbits) are interesting, since they give descriptions for 
branes that intersect with, and/or  fall on each other. The approach towards 
the origin may take place at finite or infinite values of $x$. In the former  
case, the two branes intersect or collide (depending on $\ve=+1$ or $-1$) and
pass through each other, while in the latter case they meet (and coincide) 
only asymptotically. For a quantitative description, we consider space- and 
time-like orbits separately.\\   

{\it Space-like case:}\\
The only space-like orbit passing the origin occurs for $M=1\ ,\ a>0$. In such
a case (having naturally $l=0$), we obtain a 1-dimensional periodic orbit 
which, several times with a period $L$, intersects the $x$-axis and then goes  
through a maximum at $r=r_0$. It is easy to see that in this case $U(r)$ has 
the form 
\be
U(r)=U_0{{1-r/r_0}\over{(1-r/r_1)^2}}
\label{44}\ee
where $r_0\= -(2\eta +E)/EQ \ ,\ r_1\=-(\eta +E)/EQ\ (r_0>r_1)$ and $U_0$  is 
defined as in the first row of eq.(\ref{35}). The period $L$ of this orbit 
then is calculated as
\footnote{The factor 4 in eq.(\ref{45}) is related to the fact that
the section curve of the the P-brane $Y^1=Y^1(x)$ can not involve a
sharp turning point at $Y^1=0$, but it contains two symmetric
semi-cycles with their turning points at $Y^1=\pm r_0$, since otherwise
a $\d$-force at $r=0$ would be required.}
\be 
L=4{\int}_0^{r_0}{{dr}\over{\sqrt{U(r)}}}={8\over{3Q}}(-\eta E)^{-3/2}
[4-(1-\eta E)(2+\eta E)^{1/2}]
\label{45}\ee
(note that $0<-\eta E<2$, by $a>0$). The periodicity of the orbit along the  
$x$-direction hints that $x$ may be a compactified coordinate, wrapped 
along a circle of radius $L/2\pi$, in which case the energy (stress) required  
for such a wrapping of the P-brane 
around the compact direction of the S-brane 
is obtained by reversing the eq.(\ref{45}) for $E=E(L)$.\\

{\it Time-like case:}\\
The origin crossing time-like orbits occur in each of the following 
situations:\\
(1)$\ M\geq 5$, arbitrary $E\& l$\\
(2)$\ M=4,\ l^2\leq {l_0}^2$, arbitrary $E$\\
(3)$\ M=3,\ l=0,\ $, arbitrary $E$\\
(4)$\ M=2,\ l=0,\ $, arbitrary $E$\\
(5)$\ M=1\ (l=0),\ a(E)\leq 0$.\\
As was pointed above, the orbit may pass through the origin at a finite time
in which case its passage takes place several times with a definite period
$T$, 
\be
T=2{\int}_0^{r_0} {{dr}\over{\sqrt{U(r)}}},
\label{46}\ee
or it may tend to the origin during an infinite `fall time' which is the 
characteristic of a `black' singularity at $r=0$ \cite{3}. Generally, 
the criterion for finite (infinite)-ness of the fall time is that the
contribution to the above integral from the region $r\ra 0$,
\be
\D T(r)\= {\int}_{r\ra 0}{{dr}\over{\sqrt{U(r)}}},
\label{47}\ee
vanishes (diverges). For each of the above cases, one can see that 
\bea
\left.\begin{array}{lllll}
M\geq 5 \ \ \ \ \ \ &:\ \ \ \ \ &\D T(r)\sim 1/r^{(M-4)/2}\ \ &,\ \ \ 
{\o}(r)\sim{\o}_0r^{M-4}\\ 
M=4\ \ (l^2<{l_0}^2)&: \ \ \ &\D T(r)\sim -\ln r\ \ &,\ \ \ 
{\o}(r)\sim{\o}_0\\
M=4\ \ (l^2={l_0}^2)&: \ \ \ &\D T(r)\sim 1/r\ \ &,\ \ \ 
{\o}(r)\sim{\o}_0\\
M=3\ \ (l=0)&:\ \ \  &\D T(r)\sim r^{1/2}\ \ &,\ \ \ 
{\o}(r)\equiv 0\\
M=2\ \ (l=0)&:\ \ \  &\D T(r)\sim \int{{dr}\over{\sqrt{-\ln r}}}\ \ &,\ \ \ 
{\o}(r)\equiv 0\\
M=1\ \ (l=0)&:\ \ \ &\D T(r)\sim r\ \ &,\ \ \ 
{\o}(r)\equiv 0
\end{array}\right.
\label{48}\eea
where ${l_0}^2\= E^2Q,\ {\o}_0\= l/EQ$. We have appended to this list a third
column for angular velocities near $r=0$, so as to clarify the status of the
angular motion in this vicinity. From this list it is evident that all of the 
$M\geq 4$ orbits have infinite fall times ($\D T\ra\infty$). However, while
all $M\geq 5$ orbits tend {\it linearly} (as $\o\ra 0$)
towards the origin, the $M=4$ orbits 
approach {\it spirally} to that point (as $\o\ra{\o}_0$). On the other hand,
all $M\leq 3$ orbits tending to the origin have a finite fall time 
($\D T\ra 0$), and the corresponding motions are in the form of periodic 
1-dimensional oscillations  around the origin (as $\o\equiv 0$).

\section{The S-P system with $d_s>d_p$}
Here the second row of (\ref{9}) defines $(x^a,y^A)$. The Lagrangian 
describing the mono-dimensional solutions: $Y^i=Y^i(x),\ Y^m=Y^m(x)$, with $x$  
being one of the $x^a$'s, turns to be 
\be
L=-H^{-m/2}[1+\ve ({\dot Y}^i)^2+\ve H({\dot Y}^m)^2]^{1/2}
\label{49}\ee
where the exponent $m$ is a function of dimensions as 
\be
m\= 2{\a}(d_s){\a}(d_p)+{\dt}_sd_p/\Db
\label{50}\ee
with $\a (d)$ defined in eq.(\ref{a}).
Note that in eq.(\ref{49}), $H$ is a function only of $Y^m$'s through
their modulus. As eq.(\ref{50}) shows, $m$ is a generally non-integer real
number. However, according to the `intersection rule' of the `marginal'
brane intersections \cite{1,11},
a pair of $(d_s-1,d_p-1)$-branes (not necessarily
a S-P configuration) sharing in a number $(\d -1)$ of their
worldvolume directions constitutes a marginally stable configuration provided 
\be
\d =-2{\a}(d_s){\a}(d_p)+d_sd_p/\Db
\label{51}\ee
Combining eqs.(\ref{50}),(\ref{51}), one obtains 
\be
m=d_p-\d
\label{52}\ee
showing that in such occasions $m$ is a positive integer specifying the number
of the (rotation) angles between the two branes, when they are `flattened' in  
all of their worldvolume directions \cite{1}. For an arbitrary situation, 
however one can see, using the relations of the 
type of eq.(\ref{a}), that $m$ 
satisfies 
\be
-2\leq m\leq 2
\label{53}\ee
For a marginal intersection, thus the only possibilities are $m=0,1,2$. \\
The conserved quantities (charges) corresponding to the symmetries 
(\ref{5})-(\ref{7}) consist of
\bea
&&E\={\dot Y}^i{{\p L}\over{\p{\dot Y}^i}}+
{\dot Y}^m{{\p L}\over{\p{\dot Y}^m}}-L=
{{H^{-m/2}}\over{\sqrt{1+\ve ({\dot Y}^i)^2+\ve H({\dot Y}^m)^2}}}
\label{54}\\
&&J^{mn}\= Y^m{{\p L}\over{\p{\dot Y}^n}}-Y^n{{\p L}\over{\p{\dot Y}^m}}=
-\ve{{ H^{-m/2+1}(Y^m{\dot Y}^n-Y^n{\dot Y}^m) }\over
{\sqrt{1+\ve ({\dot Y}^i)^2+\ve H({\dot Y}^m)^2}}}
\label{55}\\
&&P^i\= {{\p L}\over{\p{\dot Y}^i}}=-\ve E{\dot Y}^i
\label{56}\\
&&J^{ij}\= Y^i{{\p L}\over{\p{\dot Y}^j}}-Y^j{{\p L}\over{\p{\dot Y}^i}}=
P^iY^j-P^jY^i
\label{57}\eea
By eqs.(\ref{54}),(\ref{56}) the `velocity' components in the
$y^i$-directions are conserved quantities.
Owing to the Lorenz invariance of the action (\ref{49}) along the $(x,y^i)$
directions, one can rotate these coordinates so as to obtain ${\dot Y}^i=0$. 
In such a Lorenz frame, $Y^i$'s become constant coordinates, which can be
taken to be zero, and hence by eqs.(\ref{56}),(\ref{57}) one obtains
$P^i=J^{ij}=0$. Accordingly, the orbit $Y^m=Y^m(x),\ Y^i(x)=0 $ is described
by means of the two conservation laws (\ref{54}),(\ref{55}) now with
${\dot Y}^i=0$. Just as in the case with $d_s=d_p$, one can see that the orbit
lies in a 2D plane spanned within the $y^m$-directions. Choosing the polar
coordinates $(r,\th )$ on this plane, in analogy to eqs.(\ref{21}),(\ref{22}),
one obtains
\bea
&&{{H^{-m/2}}\over{\sqrt{1+\ve H(r)({\dot r}^2+r^2{\dot{\th}}^2)}}}=E
\label{58}\\
&&{{H^{-m/2+1}(r)r^2\dot{\th}}\over
{\sqrt{1+\ve H(r)({\dot r}^2+r^2{\dot{\th}}^2)}}}=l
\label{59}\eea
where $r^2\= Y^mY^m$. In analogy to eqs.(\ref{23}),(\ref{24}), these two
equations give
\bea
&&{\dot r}^2=U(r)\={{f(r)-l^2}\over{E^2r^2H^2(r)}}
\label{60}\\
&&{\dot\th}=\o (r)\={{l}\over{Er^2H(r)}}
\label{61}\eea
Here $f(r)$ has the following definition
\be
f(r)\=\ve r^2H(r)[H^{-m}(r)-E^2]
\label{62}\ee
It is convenient to re-express $f(r)$ as a function of $H$ as
\bea
f(H)=\left\{\begin{array}{ll}
     \ve Q^n{{H(H^{-m}-E^2)}\over{(H-1)^n}}\ \ \ &,\ \ M\neq 2\\
     \ve e^{-2H/Q}H(H^{-m}-E^2)\ \ \ &,\ \ M=2
     \end{array}\right.
\label{63}\eea
where $n\= 2/(M-2)$. Again the solutions for $(r(x),\th (x))$ are found
by integrations as in eqs.(\ref{25}),(\ref{26}) and the solution for
$r(\th )$ is found directly from eq.(\ref{27}).  
Also, as eqs.(\ref{60}),(\ref{61}) indicate, the orbital `motion' is composed
of a {\it monotonic} angular `motion' 
around $r=0$ and a radial `motion' within
a range specified by the inequality (\ref{31}). Accordingly, the
qualitative description of the orbits, as in the past, proceeds via the study 
of the $f(r)$ diagrams sketched in Figs.(2)-(6).
By eq.(\ref{62}), a marginal configuration of two parallel branes in this
case corresponds to the values $m=0,\ E=1,\ l=0$.

\subsection{An overall look on the orbits }
A large variety of the regimes and phases of the effective radial motion
are revealed, when one travels through several transverse dimensions and
changes the (the ranges of) the three parameters $(m,E^2,l^2)$.
The corresponding diagrams for the function $f(r)$
 are shown in Figs.(2)-(6) for the
case of the space-like orbits; the time-like 
case is covered by reversing the same diagrams. 
As in the previous ($d_s=d_p$) case, all the dimensions $M\geq 5$
with fixed $(m,E)$ share a common qualitative behavior. Different behaviors
appear for $M\leq4$. Some more or less novel features
in this case are the appearance of the bound, circular (stable/unstable),
origin-crossing and infinite orbits as well as the co-existing phases of the
bound and infinite orbits in the dimensions $M\geq 1$ and in both the space-
and time-like cases. A completely new behavior occurs for the case of the
space-like orbits in dimension $M=3$, where $f(r)$
possesses a pair of local minimum and maximum points.
In such a case (for suitable
values of $l^2$ ) a regime with
co-existing phases of an origin-crossing and a
bound oscillatory orbits is realized. Marginally stable circular orbits, like
the previous case, occur just for a particular time-like case in $M=4$.

\subsection{Some quantitative features}
\subsubsection{The infinite orbits}
The cases with infinitely extended orbits correspond to those values 
of $(M,m,l,E,\ve )$ that the solution to the inequality $U(r)\geq 0$ 
includes the $r\ra\infty$ region. A variety of these orbits (for all 
$M\neq 2$) are distinguishable among the Figs.(2)-(6). It is easy to 
check, as in the $d_s=d_p$ case, that all such orbits also in this case 
exhibit a straight line-like asymptotic behavior which is governed by the 
eqs.(\ref{34}), where the parameter $U_0$ is
\bea 
U_0=\left\{ \begin{array}{cll}
\ve (1-E^2)/E^2\ \ &,&\ \ M\geq 3\\
\infty\ \ &,&\ \ M=1,\; m<-1\\
1/E^2\ \ &,&\ \ M=1,\; m=-1\\
0\ \ &,&\ \ M=1,\; m>-1
\end{array}\right.
\label{64}\eea
The allowed cases are those with $U_0\geq 0$. So, using eq.(\ref{36}),
one obtains
\bea
E=\left\{\begin{array}{lll}
|cos \c | \ \ &,&\ \ \ve =+1\\
1/\sqrt{1-v^2}\ \ &,&\ \ \ve =-1
\end{array}\right. \ \ \ (M\geq 3)
\label{65}\eea
Thus, for such orbits, in the space-like case we have $ 0\leq E\leq 1$,
while in the time-like case $E\geq 1$. In the former case the $E=0$
($E=1$) bound corresponds to an asymptotic orthogonal (parallel)
configuration, while in the latter the $E=1$ ($E=\infty$) bound corresponds
to an asymptotic static (light-like) motion. 
Again, the dimension $M=1$ exhibits an exceptional behavior:
for $m<-1$ ($m>-1$) the (S,P)-branes are asymptotically perpendicular
(parallel) to each other, while for $m=-1$ they make an asymptotic angle
$\c =tan^{-1}(1/E)$.

The $E\ra 0$ limit in the space-like case is particularly interesting
because it corresponds to a generalized {\it marginal} configuration
of two branes in which one of the branes is curved. Such a configuration may
preserve some fraction of SUSY at least in some special cases.
By eqs.(\ref{60}),(\ref{61}), one sees that
 $U,\o \ra \infty$ when $E\ra 0$,  showing
that the P-brane in this limit is totally projected into a {\it planar}
curve lieing in the transverse space of the S-brane and so the two branes
are `orthogonal' to each other. One can show that the orbit's equation
(eq.(\ref{27})) in the special case with $m=1$ and in the limit $E\ra 0$ 
has the solution: $r=l/cos\th$ corresponding to an straight line of separation
$r_0=l$ relative to the source at $r=0$. This corresponds to the marginal
configuration of two different branes at one {\it right} angle preserving
a 1/4 SUSY fraction, which is a special case of
the solutions found in \cite{11,1}.

\subsubsection{The circular orbits}
Despite the $d_s=d_p$ case, here circular orbits of both the space- and
time-like natures are allowable for all $M\geq 2$. Several such orbits
are distinguished by the extremum points of the graphs in Figs.(2)-(6) 
and are classified in the Table (1) below. The symbols $s$, $u$ and $ms$
in this table refer to the stable, unstable and marginally stable orbits
respectively.

$$
\begin{array}{|c|l|l|l|l|}\hline
 &\ \ \ M=2&\ \ \ M=3&\ \ \ M=4&\ \ \ M\geq 5\\ \hline
m<-1&\ve =\pm 1(s)&\ve =+1(s,u)&\ve =+1(u)&\ve =+1(u)\\ \hline
m=-1&\ts&\ve =+1(s)&\ts&\ts \\ \hline
-1<m<0&\ts&\ts&\ts&\ts\\ \hline
m=0&\ts&\ \ \ \ \ \ -&\ \ \ \ \ \ -&\ve =\pm 1(u)\\ \hline
0<m<1&\ts&\ve =-1(s)&\ve =-1(u)&\ve =-1(u)\\ \hline
m=1&\ve =-1(s)&\ts&\ve =-1(ms)&\ts\\ \hline
m>1&\ts&\ts&\ve=-1(s)&\ts\\ \hline
\end{array} 
$$                               
\begin{center}  Table (1): circular orbits for $d_s>d_p$  \end{center}

Each orbit in Table (1) occurs only for a specific range of 
$E$ and a suitable value of $l$. The radius $R$ of a circular orbit of
energy $E$ is determined, through its corresponding value of $H$, by the 
equation
\bea
(1-m-n)H^{1-m}-(1-m)H^{-m}+E^2(n-1)H+E^2=0\ \ &,&\ \ M\neq 2\nn\\
({{2H}\over Q})^{1-m}-(1-m)({{2H}\over Q})^{-m}
-E^2({Q\over 2})^m({{2H}\over Q}-1)=0\ \ &,&\ \ M=2
\label{66}\eea

Two distinguished behaviors occur in $M=3,4$ cases. For $M=3,\ m<-1$, and at 
$E=E_m$, $H=H_m$ with
\bea
&&H_m=\sqrt{{{m-1}\over{m+1}}}\nn\\
&&E_m^2=\left ({{m-1}\over{m+1}}\right )^{-(m+1)/2}(m-1)
(m+\sqrt{m^2-1})
\label{67}\eea
the first and second derivatives of $f(H)$ simultaneously vanish and a 
`semi-stable' (space-like) circular orbit of radius $R=Q/(H_m-1)$ is 
obtained. In the range of energy $1<E<E_m$, 
this orbit splits into a pair of stable and unstable circles at different
radii at the same value of $E$ but different values of $l^2$.
For $M=4$, $m=1$, and at values $E=1, \ l^2=Q$ ,
a marginally stable (time-like) circular orbit of arbitrary
radius is realized. For the same value of $E$ but with $l^2<Q$, this circle 
is deformed into a `spiral curve' of the form 
\be
r(\th )=Re^{\n\th }
\label{68}\ee
where $\n :=\sqrt{Q-l^2}/l$ and $R$ is any arbitrary initial radius.

\subsubsection{The coincident (S,P) configurations}
In addition to the above mentioned finite size circular orbits, there 
exist other limiting cases which are characterized by circles of zero 
radius with a vanishing, finite or infinite angular velocity. 
In a time-(space-)like situation, such a solution describes
a  coincident (S,P) configuration in which the P-brane, in some cases, 
has a non-vanishing spin (torsion) equal to $\o (r\ra 0)$. 
Such orbits, which are the $r=0$ solutions to the equations of motion, 
are specified by those values of the parameters satisfying
\be
U(r\ra 0)=0\ \ ,\ \ U'(r\ra 0)=0
\label{69}\ee
The solution is stable (unstable) provided $U''(r\ra 0)$ to be negative
(positive). The expansion of $U(r)$ near $r=0$ up to a few leading terms
for different values of $(M,m)$, thus gives rise to a classification of 
the coincident (S,P) solutions as is indicated in Table (2) below.

$$
\begin{array}{|c|c|c|}\hline
  M=3&M=4&M\geq 5\\
  (\o =\infty)&(\o ={\o}_0)&(\o =0)\\ \hline

m=-1&-{1\over 2}<m<0&{{3-M}\over{M-2}}<m<0\\
\ve =+1(s)&\ve =-1(s),+1(u)&\ve =-1(s),+1(u) \\ \hline
-&m=0&m=0\\
&\ve =\pm 1(s,u)&\ve =\pm 1(s,u)\\ \hline
-&0<m<1&m>0\\
&\ve =\pm 1(s)&\ve =+1(s),-1(u)\\ \hline
-&m\geq 1&-\\
&\ve =\pm 1(s),-1(u)&\\ \hline
\end{array} 
$$                               
\begin{center}  Table (2): coincident (S,P) 
solutions for $d_s>d_p$  \end{center}

As in Table (1), each of the solutions in Table (2) is achieved for a 
suitable range or fixed values of $(E,l)$. As the first row in this table
shows, while the $M=4$ solutions carry a finite spin or torsion ($\o ={\o}_0 
:=l/EQ$), all the $M\geq 5$ solutions are without spin and torsion ($\o =0$).
On the other hand the $M=3$ solution has an infinite torsion ($\o =\infty$)
which means that one of the P-brane directions is crumpled (i.e.,wrapped
on a circle of vanishing radius) around the S-brane directions. 

\subsubsection{The origin-crossing orbits}
Orbits of this type occur in cases where the solution to the inequality
$U(r)\geq 0$ involves the $r=0$ point. Generically, such an orbit is bounded
between $r=0$ and $r=r_0$, where $U(r_0)=0$ (through in some cases 
$r_0\ra \infty$). The origin may be crossed or approached by the orbit 
depending on that the quantity defined in eq.(\ref{47}) (which is denoted
by $\D X$ below to include both the space- and time-like cases) vanishes
or diverges respectively. The various origin crossing orbits of 
both characters, for different values of $(M,m)$, are classified in Table (3) 
below.
$$
\begin{array}{|c|c|c|c|c|c|}\hline
(\ve ,\D X)&M=1 & M=2& M=3& M=4& M\geq 5\\ \hline
m<0&(\pm 1 ,0)&(+1,0)&(+1,0)&(+1,0)&(+1,0)\ ,\ m<{{4-M}\over{M-2}}\\ 
& & & & &(+1,\infty )\ ,\ m\geq{{4-M}\over{M-2}}\\ \hline
m=0&''&(\pm 1,0)&(\pm 1,0)&(\pm 1,\infty )&(\pm 1,\infty )\\ \hline
m>0&''&(-1,0)&(-1,0)&(-1,\infty )&(-1,\infty ) \\ \hline
\end{array} 
$$                               
\begin{center}  Table (3): origin-crossing orbits for $d_s>d_p$  \end{center}

\section{The S-P system with $d_s<d_p$}
By defining $(x^a,y^A)$ as in the third row of (\ref{9}), the Lagrangian $\cL$
describing the embedding coordinates $Y^A=Y^A(x^{\m},y^i)$ is found to be
\be
\cL =-H^{-m/2}det^{1/2}({\d}^{\m}_{\n}+H{\p}^{\m}Y^A{\p}_{\n}Y^A)
det^{1/2}({\d}_{ij}+{\p}_{i}Y^A{\p}_{j}Y^A)
\label{101}\ee
where $m$ is defined as in eq.(\ref{50}) with $(d_s,d_p)$ interchanged, i.e.
\be
m\= 2{\a}(d_s){\a}(d_p)+d_s{\dt}_p/\Db
\label{102}\ee
When the `marginality' condition (\ref{51}) is satisfied, in analogy to
eq.(\ref{52}), $m$ coincides to the 
number of the non-zero angles for marginal intersection:
\be
m=d_s-\d
\label{103}\ee
A major difference between this ($d_s<d_p$) case and the previous
($d_s\geq d_p$) ones is that the Lagrangian (\ref{101}), unlike
(\ref{12}),(\ref{49}), contains explicit dependences on (the subset $y^i$ of)
the parameterization coordinates $(x^{\m},y^i)$ themselves, through the
dependence of $H$ on
\be
r=\sqrt{(y^i)^2+(y^A)^2}
\label{104}\ee
This property (generally) prevents the 
possibility of existing any mono-dimensional
solution, which as in the 
previous cases depends only on one of the coordinates
$(x^{\m},y^i)$. In particular, parallel relatively moving configurations in
this case are not allowable (this is unlike to the situation where the roles
of $d_s$ and $d_p$ are interchanged !). However, other interesting possibilities
for solutions of the Lagrangian (\ref{101}) exist which are explained below.\\

{\it (Semi-)cylindrical configurations:}\\
Interesting configurations occur when a number $\d$ of the {\it flat} S-brane
directions are parallel, and the rest $(d_s-\d )$ of them are perpendicular,
to a (generally) {\it curved} P-brane. This means that (for each $A$) a
number $\d$ of the slopes ${\p}_{\m}Y^A$ are vanishing, while the rest
$(d_s-\d )$ of them tend to infinity
\footnote{More precisely, the slopes for perpendicular $x^{\m}$'s are in the
form of $\d$-functions at the corresponding values of these coordinates.}.
Putting these values in eq.(\ref{101}), one peaks up a factor
 $H^{(d_s-\d)/2}$ multiplied by a constant infinite factor. Thus, ignoring
 the irrelevant infinity, the dynamics of the coordinates $Y^A=Y^A(y^i)$
 in this case is governed by
 \be
{\cL}^*=-H^{(d_s-\d -m)/2}det^{1/2}({\d}_{ij}+{\p}_{i}Y^A{\p}_{j}Y^A)
\label{105}\ee
The case of a {\it totally cylindrical} configuration happens when all of the
S-brane directions are common to the P-brane, and is covered by the
above Lagrangian for $\d =d_s$.\\

{\it Curved marginal intersections:}\\
An interesting subclass of the above semi-cylindrical solutions consist
of the marginal intersections in which one of the two branes (the P-brane)
is not necessarily flat. To see that such solutions exist, imagine 
a pair of branes which are capable of forming a flat marginal solution;
that is they obey the rule (\ref{51}). Assuming a 
semi-cylindrical configuration with $\d$ cylindrical directions, thus the
Lagrangian ${\cL}^*$ becomes
\be
{\cL}^*=-det^{1/2}({\d}_{ij}+{\p}_{i}Y^A{\p}_{j}Y^A)
\label{106}\ee
which, unlike eq.(\ref{105}), does not contain any trace of the spacetime
curvature via dependence on $H$. The classical solutions for this
Lagrangian do represent $(d_p-d_s)$-dimensional $minimal$ (hyper)surfaces in a
$({\dt}_s+2)$-dimensional Euclidean space
 parametrized by $(y^i,y^A)$.
These surfaces are solutions to the Laplace equation in the curved space,
\be
{\Box}_gY^A=0
\label{107}\ee
where ${\Box}_g\={1\over{\sqrt g}}{\p}_i\sqrt g g^{ij}{\p}_j$ and
$g_{ij}[Y]\={\d}_{ij}+{\p}_{i}Y^A{\p}_{j}Y^A$ is the pull-back of the
Euclidean $({\dt}_s+2)$-dimensional metric on the minimal surface. For
asymptotic flat boundary conditions, the only solutions to this equation are
flat surfaces identifying ordinary (flat) orthogonally intersecting S-P
systems. For non-asymptotic flat surfaces or surfaces with non-planar
boundaries, however, the above equation have non-trivial solutions leading
to the curved marginal intersections. In this manner, given every such minimal
surface, we can construct a solution of the original theory (\ref{101})
simply by a cylindrical extension of 
the minimal surface in some of the S-brane directions.

\section{Conclusion}
In this paper, we have considered the `mono-dimensional'
worldvolume  solutions for one of a pair of interacting branes
in both the space- and time-like configurations, when the other brane
is kept fixed and flat (called as the P and S-branes, respectively). 
For such solutions, the S-brane, which itself has a flat
worldvolume geometry, causes the worldvolume of the P-brane to be curved
only in one of its (space- or time-like) directions. It was observed that
such a configuration is allowed {\it only} in cases where $d_s\geq d_p$.
This, for example, implies that when $d_s<d_p$, parallel (S,P) configurations
in relative transverse motion can not be stable and the motion imposes
a curvature on (at least) one of the directions of the P-brane.
In all the
allowed cases, however, it was shown that the P-brane worldvolume solutions
of this type can be described in terms of a set of planar orbits in the
subspace transverse to the S-brane, which are parametrized by their `energy'
$(E)$ and `angular momentum' $(l)$ variables. These two parameters fix the
form of the solution up to any arbitrary rotation (translation) along the
transverse (parallel) directions of the S-brane.
These orbits possess different phases, for different ranges of
the parameters $(E,l)$, corresponding to phases of an effective 2D motion.
These phases are qualitatively described in terms of a set of phase diagrams,
corresponding to different values of $(E,l)$, or equivalently by the
intersections of the graph of an $E$-dependent function $f(r)$ and a
 line of the constant height $l^2$. The `effective potential'
$-{1\over 2}U(r)$  between (S,P) and their relative `angular velocity'
$\o (r)$ as a function of the transverse distance $r$ are closely
related to this function. Though this  function has different expressions
in the cases with `similar' and `non-similar' branes, its dependence on
$(d_s,d_p,D)$ always appear through the combinations $(M,m)$ as
\bea
&&M={\dt}_s+2\nn \\
&&m=2\a (d_s)\a (d_p)+{\dt}_s d_p/(D-2)
\label{108}\eea
One of the features that the above analysis reveals is that all $M\geq 5$
cases, for the same values of $(E,l)$, share common qualitative behaviors,
while different behaviors appear in $M\leq 4$ cases.  These diagrams
also reveal several phases of the `effective motion' such
as: the bounded (oscillatory), unbounded (asymptotic free) and circular
(rotating) regimes 
 for  the space- as well as the time-like configurations.
It should be remarked that, although in the previous sections we have
considered the cases with $d_s=d_p$ and $d_s> d_p$ as representing the
configurations with two `similar' and `non-similar' branes, one can fit
also some other cases of a two brane system in the same analysis.
For example, in 10D supergravity, either of the two configurations
of (NS1,D1) and (NS5,NS1) branes can be analyzed as in the $d_s\neq d_p$
category (section 5). The reason, as can be checked,
is that in both these cases the
the form-field interaction of the
two branes (the WZ term in eq.(\ref{4}))
has a vanishing contribution to their effective action. 
Also it should be recognized that the analysis in this paper ignores
the effects of the internal gauge fields as well as the external
$B$-field on the dynamics of p-branes and treats all types of branes
as obeying similar dynamics. A possible extension
of this work thus may be obtained
by inclusion of the above fields in the branes dynamics
similar to the one which has been done recently in \cite{12}
to handle a $(D3,D5)$ -branes as a (S,P) system in another geometry.
As we have mentioned in the introduction, looking for the exact form of the
the p-branes in a multi-angle asymptotic flat intersection
has been the main motivation for writing this
 paper. Nevertheless, the solutions we have considered here
 are not suitable for describing the
 configurations with more than one angle.
Further, the one-angle asymptotic flat solutions in sections 4,5
are not of the type of the  {\it marginal} intersections
with several angles introduced
 in \cite{7,7*}. For looking for
 $m$-angle configurations, in general, one has to seek for the worldvolume
 solutions of the P-brane which depend on $m$ of the directions of the S-brane.
 Accordingly, one  has to solve a set of second order PDE's in $m$ dimensions
 which unlike the 1-dimensional case can not be integrated using its
 conservation equations. A perturbative approach in \cite{4}, however,
 shows that solutions of the marginal type for these configurations can be
 found, provided that the `marginality' property is appropriately defined
 at infinitely long distances.\\

\begin{center} {\Large Acknowledgements}\end{center}
The author would like to thank H. Arfaei for his constant encouragements
during the course of this work and careful reading of the preprint.

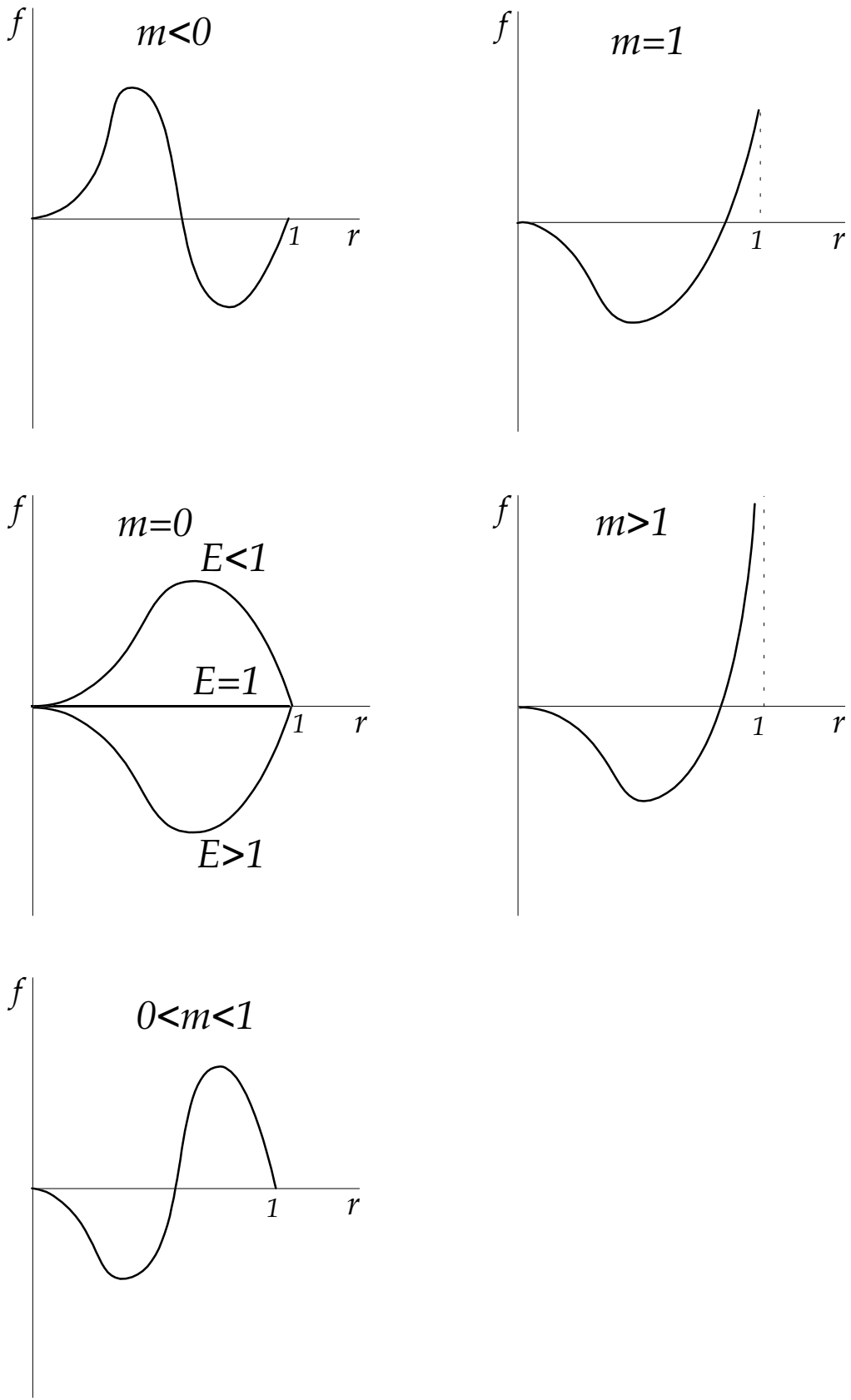
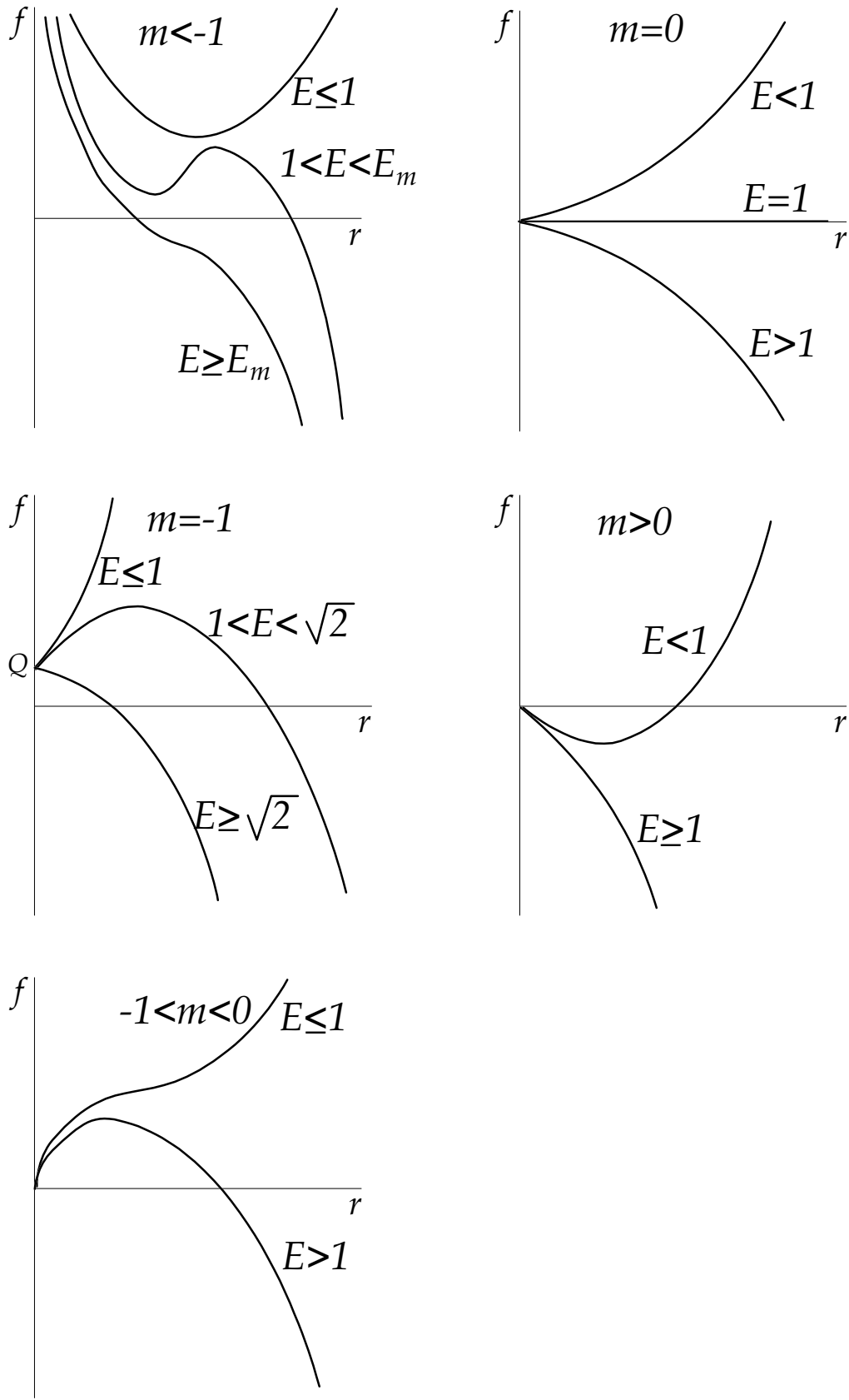
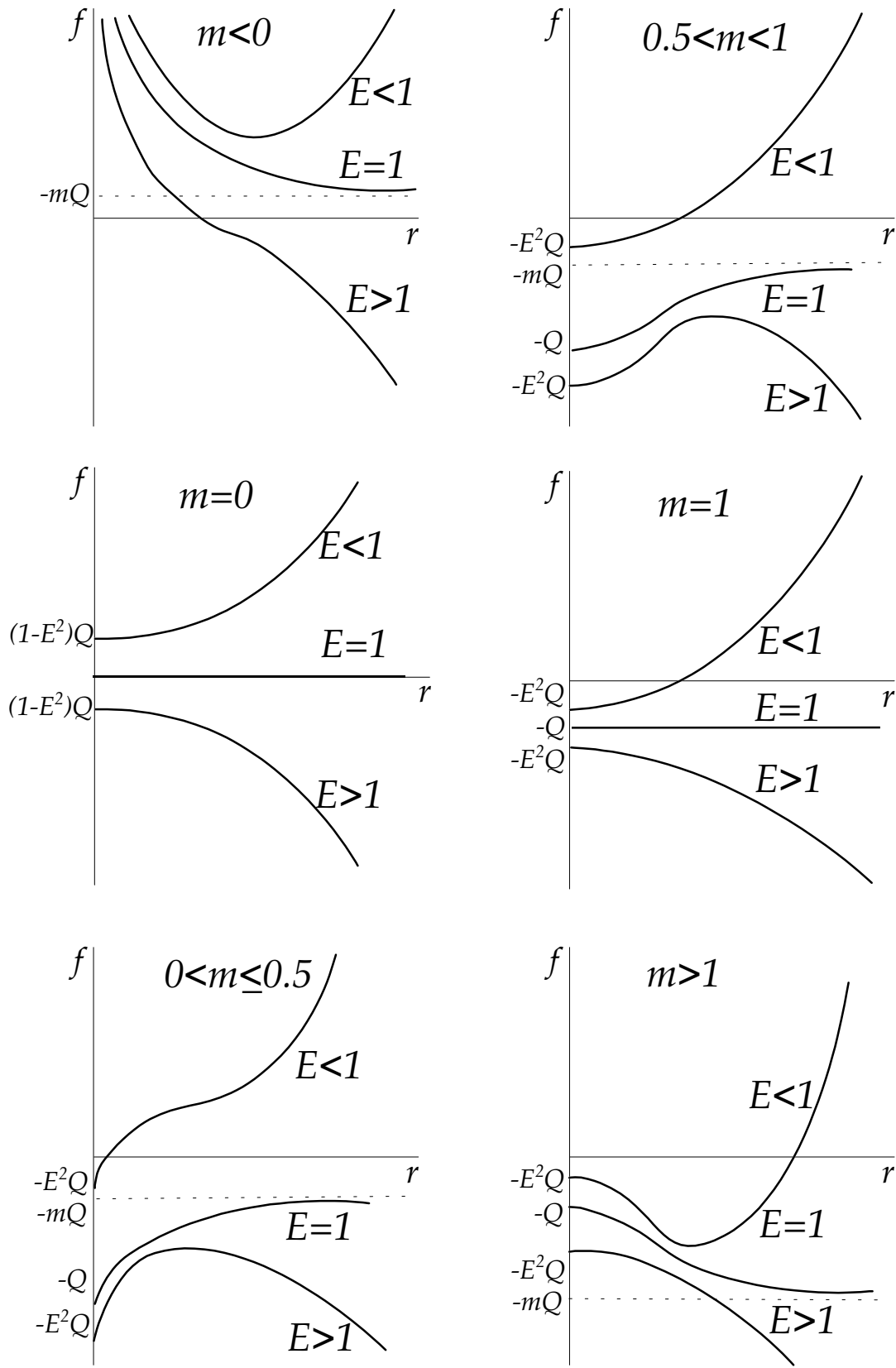
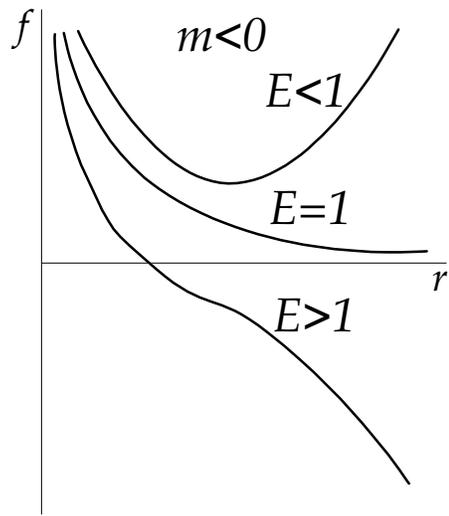
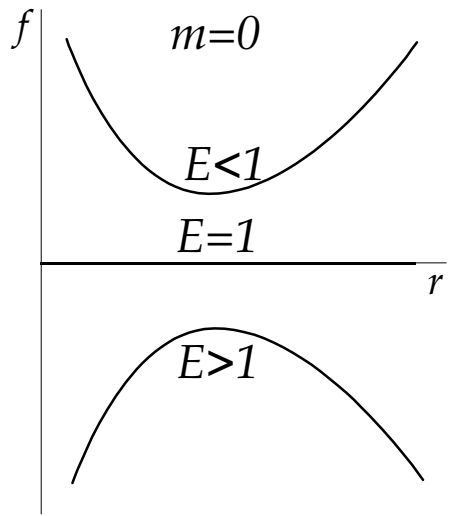
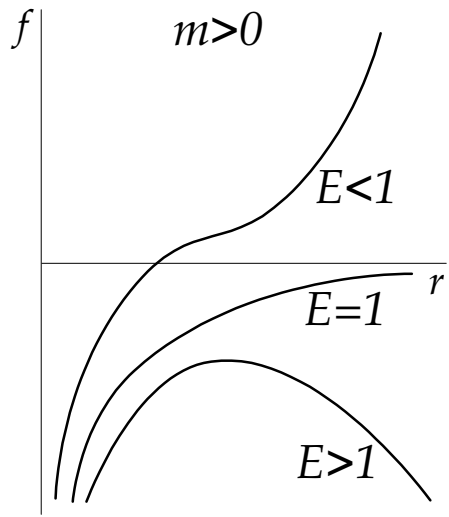
\end{document}